\documentclass[twocolumn, floatfix]{revtex4}

\usepackage{amssymb}
\usepackage{amsmath}
\usepackage{graphicx}

\begin{document}

\title{Scale Invariance in Road Networks}
\author{Vamsi Kalapala$^{\dag}$, Vishal Sanwalani$^{\dag}$, Aaron Clauset$^{\dag,}$\footnote{Corresponding author.} and Cristopher Moore$^{\dag,\ddag}$}
\affiliation{$^{\dag}$Department of Computer Science and $^{\ddag}$Department of Physics and Astronomy, University of New Mexico, Albuquerque, NM 87131. \\
{\tt (vamsi, vishal, aaron, moore)@cs.unm.edu}}

\begin{abstract}
We study the topological and geographic structure of the national road networks of the United States, England and Denmark. By transforming these networks into their {\em dual} representation, where roads are vertices and an edge connects two vertices if the corresponding roads ever intersect, we show that they exhibit both topological and geographic scale invariance. That is, we show that for sufficiently large geographic areas, the dual degree distribution follows a power law with exponent $2.2 \leq \alpha \leq 2.4$, and that journeys, regardless of their length, have a largely identical structure. To explain these properties, we introduce and analyze a simple fractal model of road placement that reproduces the observed structure, and suggests a testable connection between the scaling exponent $\alpha$ and the fractal dimensions governing the placement of roads and intersections.
\end{abstract}

\maketitle

\section{Introduction}
Complex networks has received much attention from the physics community and beyond in the recent past~\cite{RekaBarabasi02, DM02, Newman03d}. This interest has primarily sprung from the near ubiquity of networks in both the natural and manmade world. Canonical examples of complex networks include the Internet~\cite{FFF99}, the World Wide Web~\cite{Kleinberg99b}, social contacts~\cite{WF94}, scientific citations~\cite{Price65, Redner04} and gene and protein interactions~\cite{Ito01, Montoya02}. Most of these studies have focused on topological quantities like the degree distribution, diameter and clustering coefficient. More often than not, it has been found that networks exhibit a degree distribution in which the fraction of vertices with degree $k$ has the form of a power law, \mbox{$P(k) \sim k^{-\alpha}$}, where $2 < \alpha < 3$.

While virtual networks like the World Wide Web, or interaction networks like that of proteins, may be considered purely in terms of their topology, physical networks have additional geographic properties. In particular, creating and maintaining edges presumably requires physical resources proportional to their length, and the physical length of a path between two vertices may be rather different from its topological length (i.e., the number of edges along it). 
In some cases, the interaction of a network's topology with its underlying geography has been studied previously through models of evolving networks or optimizing resource costs~\cite{Newman04, FKP02, Manna03}. 

Here, we focus on the presence of hierarchy and scale invariance in physical networks as illustrated by the nationwide road networks of the United States, England and Denmark. To reveal their topological organization, we employ the {\em dual} model of the road network, in which a vertex represents a single road of a given name, and two vertices are joined if their corresponding roads ever intersect. This should not to be confused with the dual of a planar graph, in which faces become vertices and vice versa. This graph transformation has been used previously to study the topological structure of urban roads~\cite{Latora04, Jiang04, Rosvall05, Latora05}.

By representing the road network in this manner, we are able to show empirically that the degree distribution has a heavy tail, and is well-characterized by a power law with an exponent \mbox{$2.2 \leq \alpha \leq 2.4$}. Rosvall et al. showed that urban networks also have heavy tails in the dual degree distribution, although not unequivocally with a power-law form~\cite{Rosvall05}. Additionally, we find the structure of journeys on the physical network is scale invariant, i.e., the structure of a journey is similar regardless of its scale. To explain these properties, we introduce and analyze a simple fractal model for the hierarchical placement of roads on the unit square. We show that the recursive nature of this model generates the scale invariant journey structure, and suggests a simple relationship between the scaling exponent of the dual degree distribution $\alpha$ and the fractal dimensions governing the placement of roads and intersections.

\section{Primal and Dual Models}
The natural representation of a road network is a collection of road segments, in which each segment terminates at an intersection; this is called the {\em primal} representation. However, this representation gives us little opportunity to consider scale-free properties or heavy-tailed degree distributions: almost all vertices have degree $4$, and the average degree of a planar network is at most $6$, as the maximum number of edges is $3n-6$. However, this representation violates the intuitive notion that an intersection is where two roads cross, not where four roads begin. Nor does it well represent the way we tell each other how to navigate the road network~\cite{footnote}, e.g., ``stay on Main Street for $10.3$ miles, ignoring all cross-streets, until you reach Baker Street, then turn left.'' If we use the dual representation, however, such a set of directions reduces to a short path through the network where each transition from one road to another corresponds to a single step.

In order to transform the road network into its dual representation, we must define which road segments naturally belong together. In previous studies of road networks, segments have been grouped by their street name~\cite{Rosvall05, Jiang04}, line of sight by a driver~\cite{Hiller}, or by using a threshold on the angle of incidence of segments at an intersection~\cite{Latora04}. Here, we use the method of taking a single road to be the collection of road segments that bear the same street name.

\section{Sampling Methodology}
We sampled the national road networks of the United States, England and Denmark by querying a commercial service, provided by Mapquest.com. This service provides driving directions, i.e., a path through the dual graph, when given source and destination addresses. If only partial information is provided, e.g., the postal code, the service defaults to a unique address near that region's center. The directions returned are a list of road names, the respective distances a driver should travel on each, and instructions as to how to get from one road to another, e.g., ``turn left onto'' or ``continue on''.

We sampled each road network by taking the union of paths between a large number of uniformly random pairs of source and destination postal codes. For the United States, we sampled until every postal code was present in the network (roughly $200~000$ trials), while for England and Denmark, we sampled beyond this limit (for about $35~000$ total, each). We repeated our analysis for a random fraction $0.25$, $0.50$ and $0.75$ of each sample and found that our observations are not sensitive to the number of trials.

Notably, our sampled networks are biased according to population distribution, as postal codes in each country are distributed roughly according to population. 
On the other hand, by focusing only on travel between postal regions, we restrict ourselves to studying the structure of long journeys. Naturally, we expect short-range travel to represent the majority of real journeys, e.g., trips to the office, the grocery store, etc. Finally, while most details of the algorithm that Mapquest uses to generate its driving directions are concealed on account of it being proprietary, we note that any algorithm that minimizes travel time, as opposed to geographical distance, will create a bias toward traveling on major roads and highways.

\section{Journey Structure}
Intuitively, a road network is composed of a hierarchy of roads with different importance. For instance, a road atlas classifies roads, according to their speed limit and capacity, into minor and major local streets, regional roads, and finally highways. Assuming that a driver wishes to reach her destination as quickly as possible, we may model the structure of an arbitrary journey as follows. Our driver begins at the local street where her point of origin is located, and moves to progressively larger and faster roads, i.e., she moves up the hierarchy, until she reaches the fastest single road between her source and destination. On this road, she covers as much distance as possible, and then descends to progressively smaller roads until she reaches the local street of her destination.

Thus, we expect that the largest steps of a journey will cover a significant fraction of the total distance, and that the length of a step will increase as a driver moves up the hierarchy in the beginning of the journey, and decrease as she descends it at the journey's end. Empirically, we find that this assumption reflects the structure of journeys through our sampled networks. For the purposes of comparison, we classify journeys into three roughly equally populated groups based on their length: short, medium and long. 

To more precisely compare the journey structure between trips of different lengths, we define a journey's {\em profile} in the following way. We take the largest step of the journey, in terms of distance traveled, the three largest steps (in order of appearance) that precede it, and the three largest steps (again, in order of appearance) that follow it. Thus we ignore the many small steps that are scattered throughout the journey, e.g., taking a highway ramp to merge onto a national highway. While this definition of a journey profile is somewhat arbitrary, it allows us to focus on the journey's large-scale structure.

\begin{figure*}[t] 
\begin{center}
\begin{tabular*}{17.5cm}{ccc}
\hspace{0.2mm}
\includegraphics[scale=0.315]{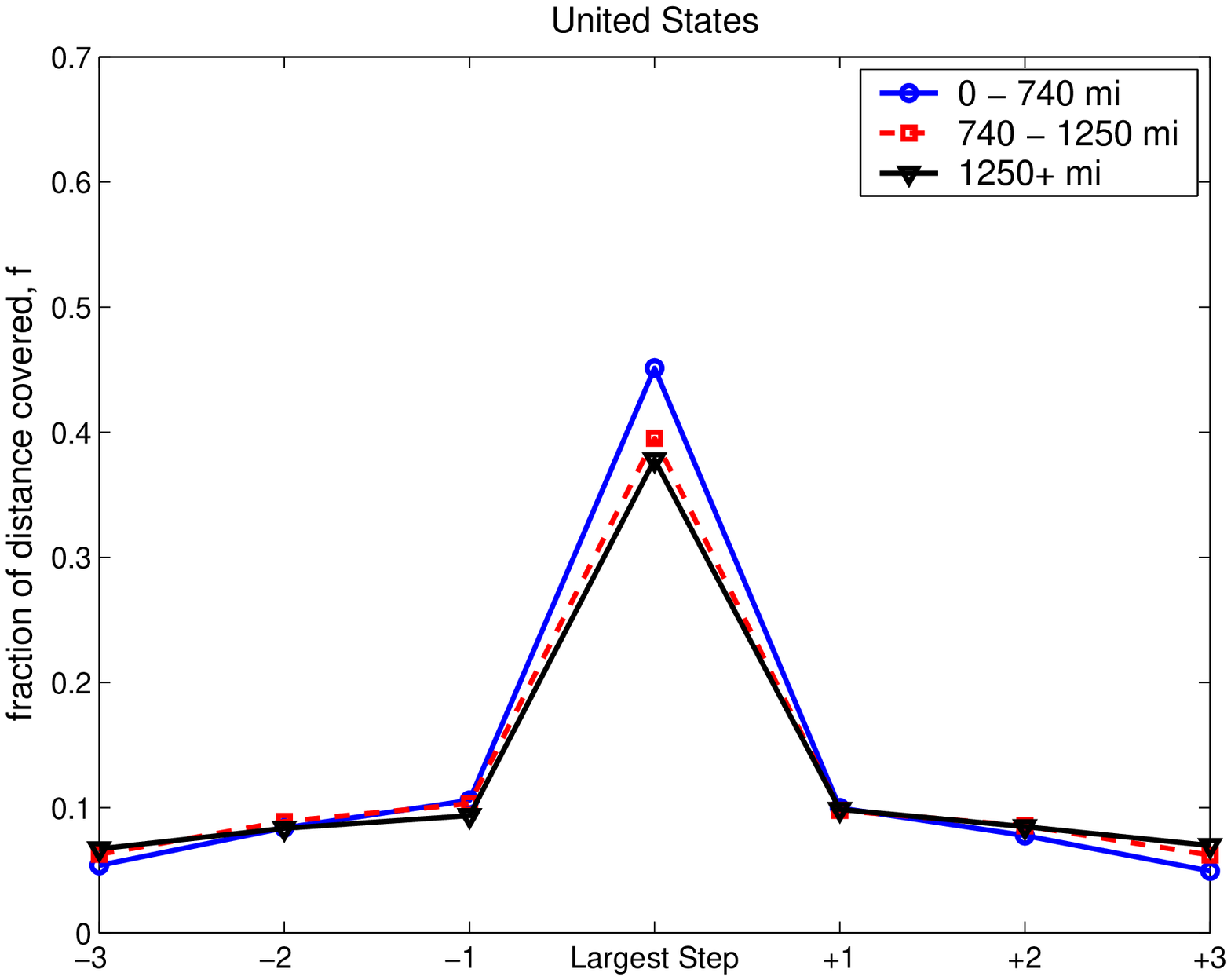} & \hspace{0.05mm}
\includegraphics[scale=0.315]{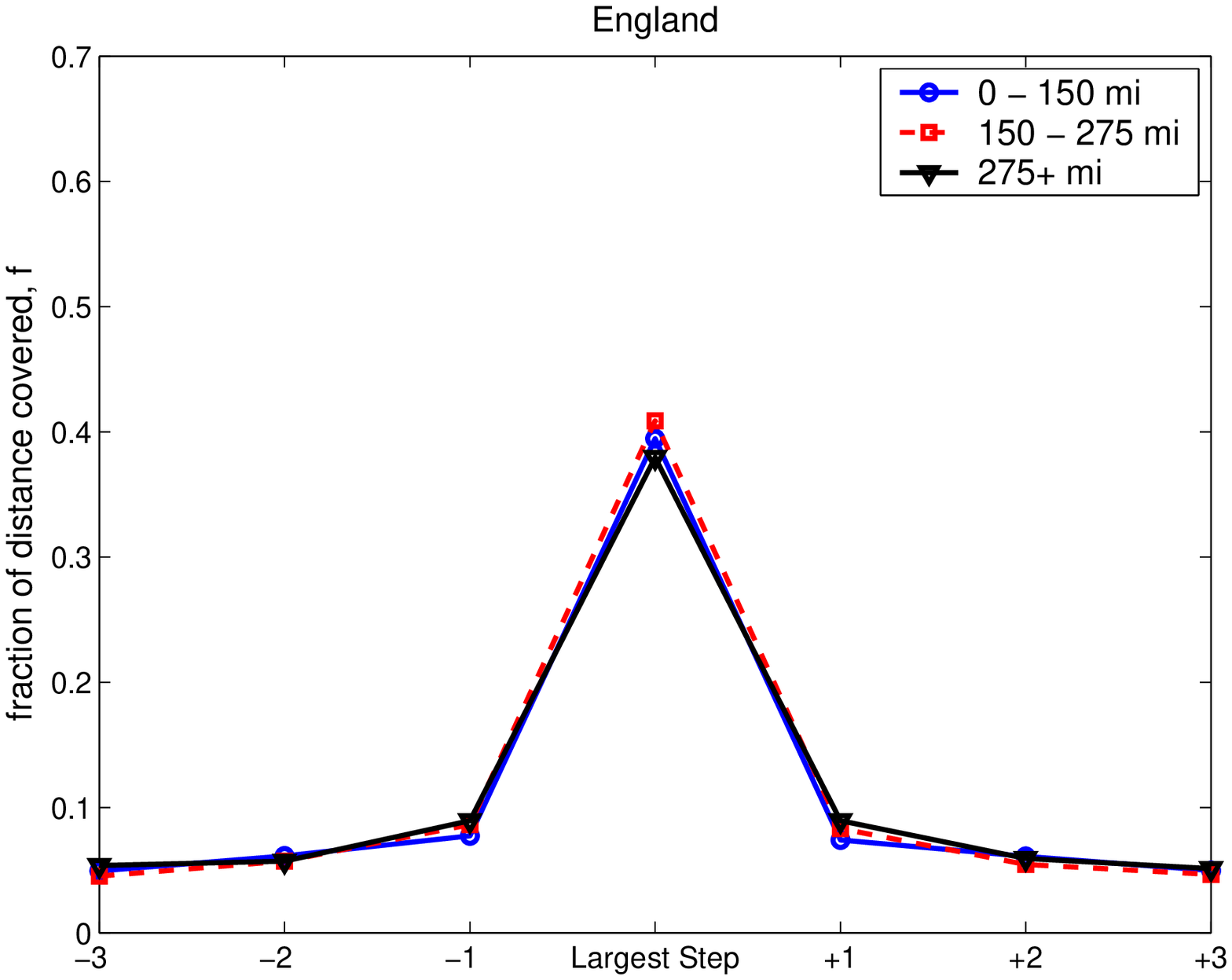} &  \hspace{0.2mm}
\includegraphics[scale=0.315]{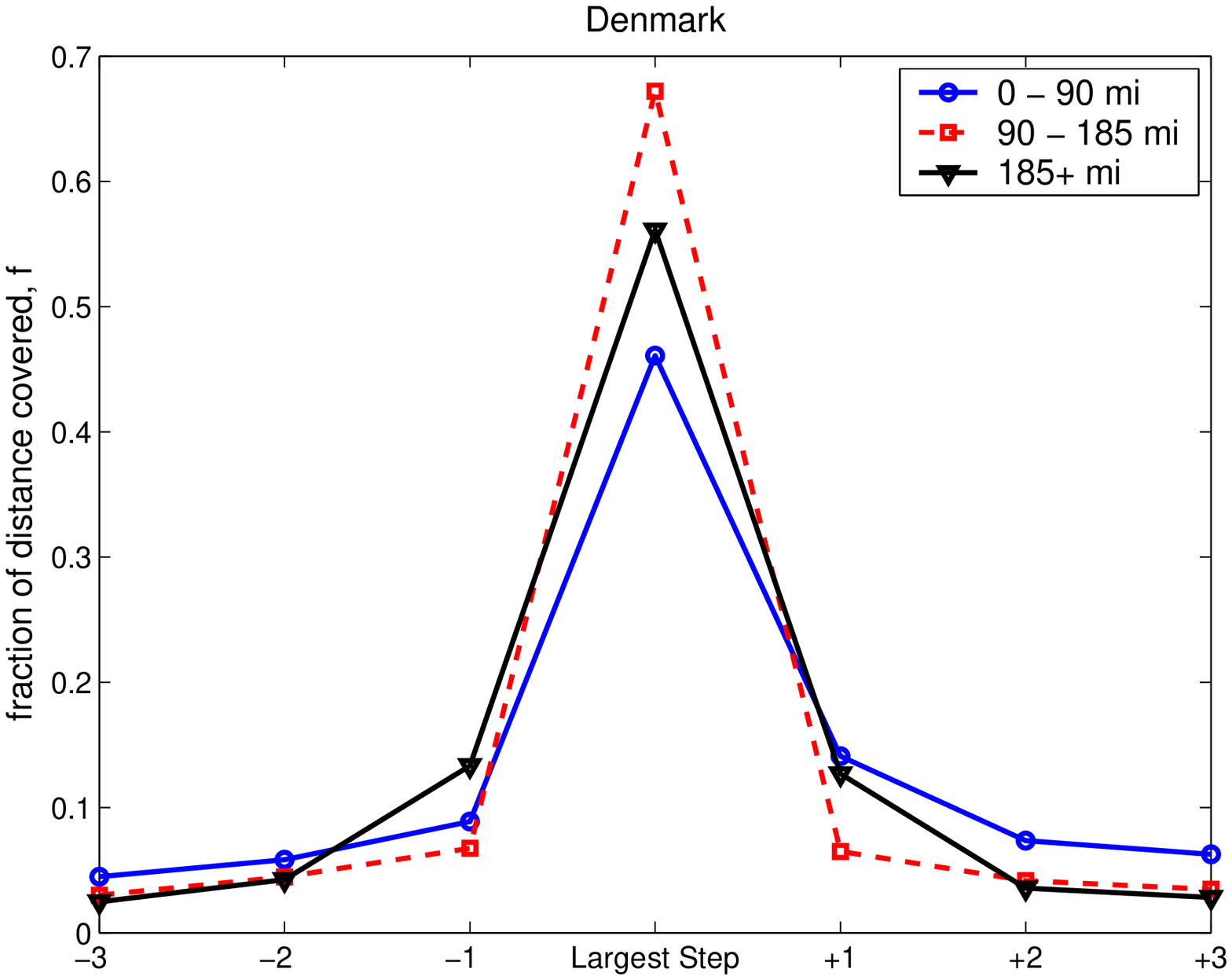} 
\end{tabular*}
\end{center}
\caption{(Color online) The average journey profiles for the United States, England and Denmark. Profiles are defined to be the largest step (centered) in the journey, flanked by the three largest preceding and subsequent steps (in order of appearance) in the path. }
\label{figure:profiles}
\end{figure*}

Figure~\ref{figure:profiles} illustrates the average profile for journeys on each of the three national road networks for short, medium and long journeys. The unimodal shape of these profiles clearly supports the hierarchical model we describe above. Additionally, their approximate collapse across journeys of different lengths indicates that the structure of the journey profile is invariant with respect to the scale of the journey. We omit error bars in Figure~\ref{figure:profiles} for visual conciseness, but note that they are less than \mbox{$\pm 0.1$}, on average. The slight asymmetry in Denmark's short-journey profile reflects the presence of one-way roads of different lengths.

\begin{figure*}[t] 
\begin{center}
\begin{tabular*}{17.5cm}{ccc}
\includegraphics[scale=0.315]{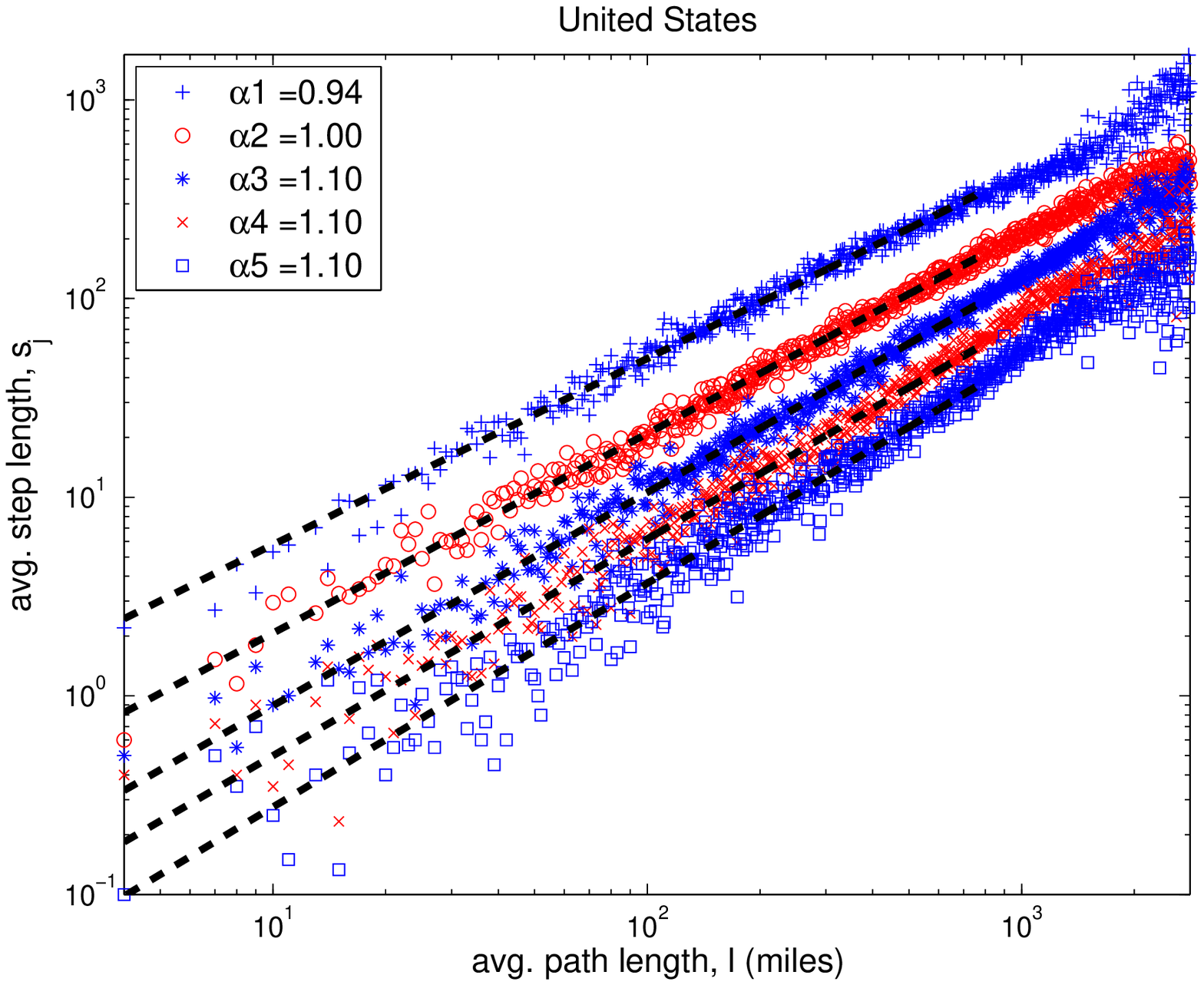} &
\includegraphics[scale=0.315]{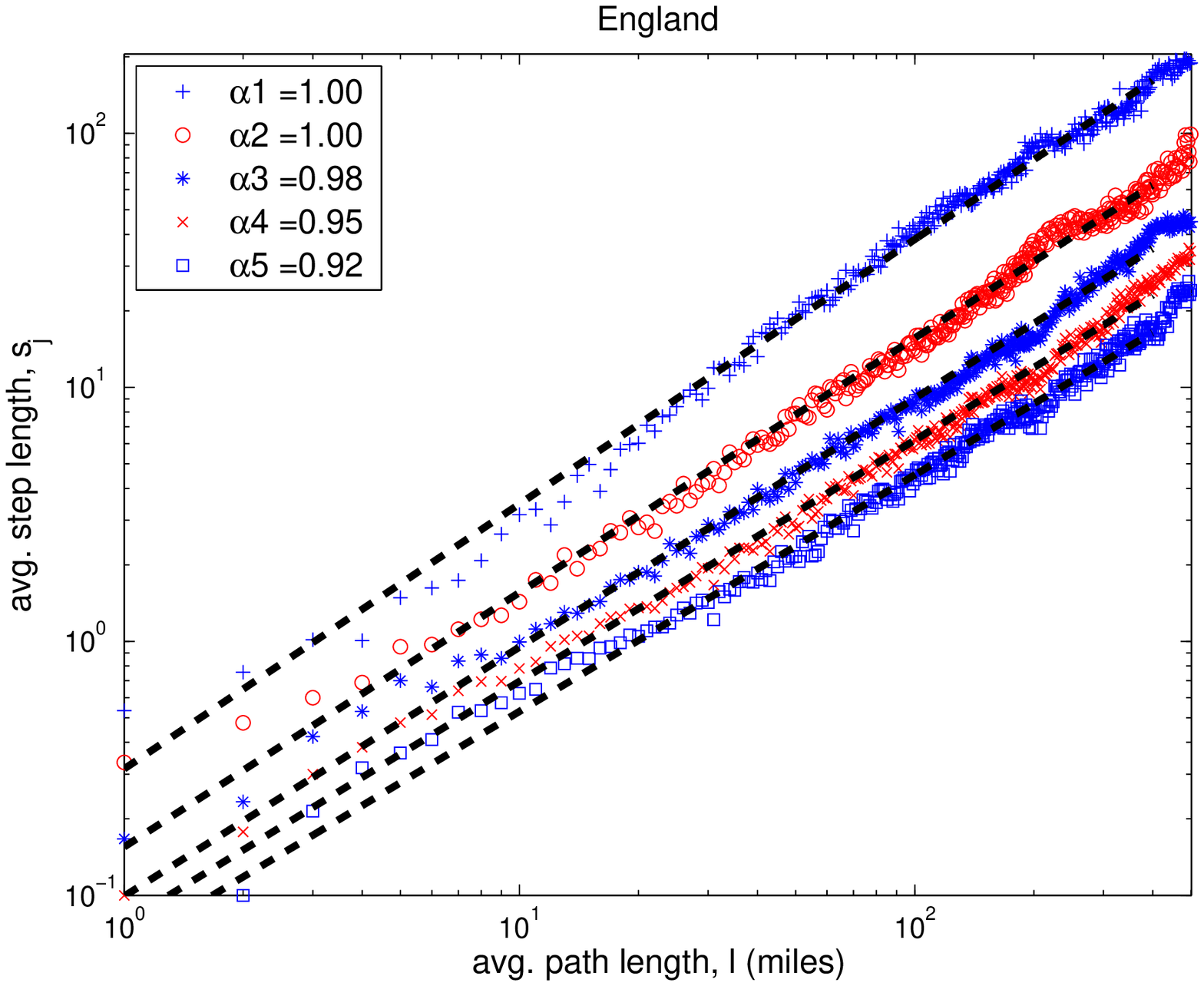} &
\includegraphics[scale=0.315]{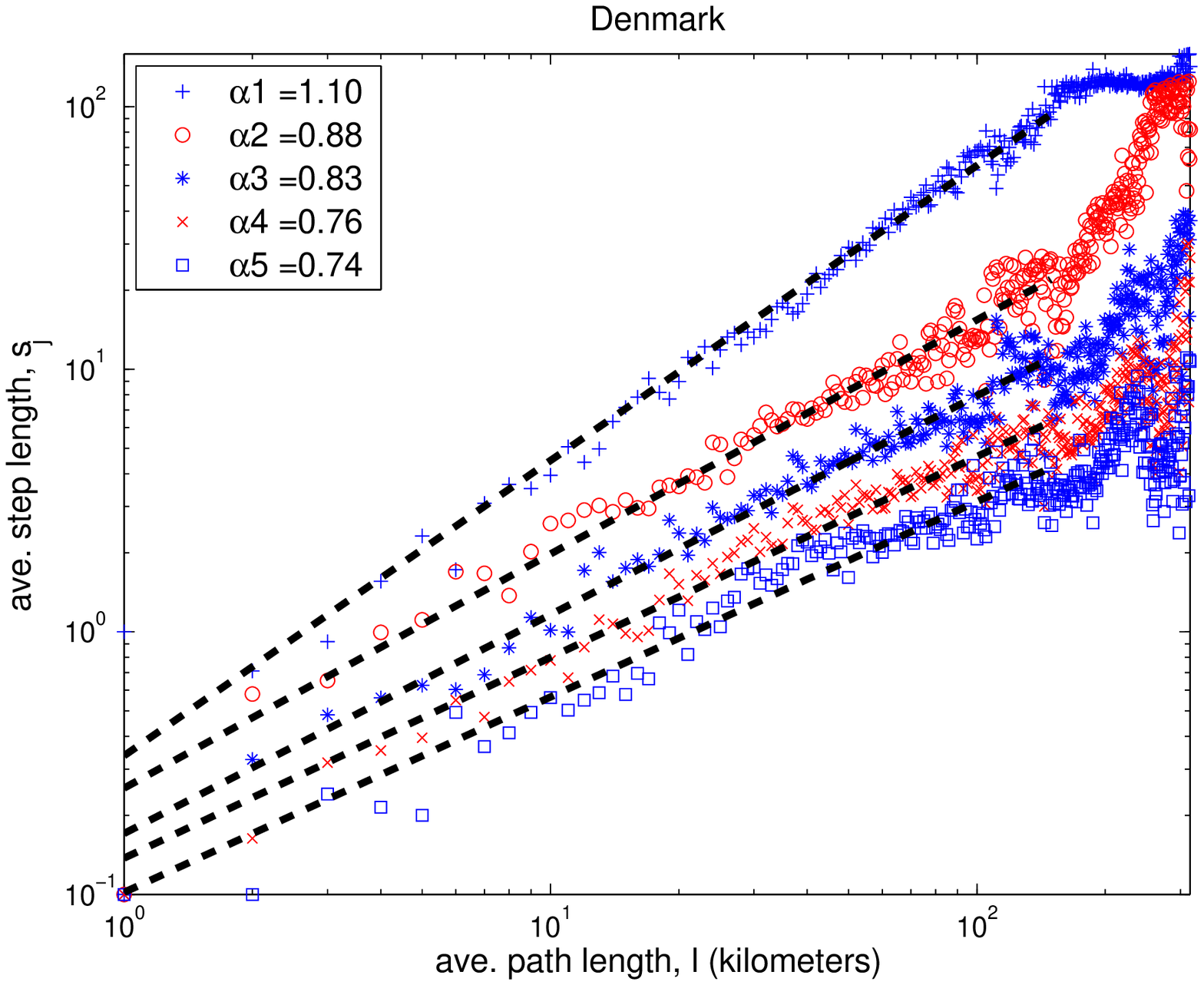} 
\end{tabular*}
\end{center}
\caption{(Color online) The scale invariant hypothesis predicts that $s_{j} \approx A_{j}\, \ell$ for constants $A_{j}$, and thus that $\alpha_{j} \approx 1$. This is consistent with our power-law fits, in which we estimate $\alpha_{j}$ using a bootstrap resampling method. Journeys on the very largest scales were excluded in order to avoid finite-size effects. }
\label{figure:steps}
\end{figure*}

In Table~\ref{table:meanSteps}, we show the fraction of the total distance covered by the five largest steps of these journeys, where $s_{j}$ is the $j$th largest step. These steps alone account for about $85\%$ of the total length of the journey; the largest step typically covers about $46\%$ of the entire distance, the second largest covers $19\%$, the third largest covers $10\%$, etc. Moreover, for each $j$ from $1$ to $5$, the fraction of the journey covered by the $j$th largest step appears to be roughly constant. This suggests a simple linear relationship of the form
\begin{equation}
\label{eq:sk}
s_{j} = A_{j}\, \ell \enspace ,
\end{equation}
where $s_{j}$ is the $j$th largest step, $\ell$ is the total path length and $A_{j}$ is some constant. Figure~\ref{figure:steps} shows the average step size for each of the five largest steps against the total path length for each of our three networks. We fit our data to a power-law with the form $s_{j} = A_{j}\, \ell^{\,\alpha_{j}}$, bootstrapped via least-squares (we ignore the longest journeys, since we expect finite-size effects to appear as $\ell$ approaches the diameter of the country). We observe that this power law fits the data reasonably well, with average $r^{2}$ values of $0.97$, $0.99$ and $0.99$ respectively; moreover, averaging across all such models, we have $\alpha_{j}= 1.0\pm 0.1$, suggesting that the linear form of~\eqref{eq:sk} is accurate.

\begin{table*}[t]
\begin{ruledtabular}
\begin{tabular}{l|ccc|cccccc|c}
Network & Distance & & & $1^{st}$ Largest & $2^{nd}$ Largest & $3^{rd}$ Largest & $4^{th}$ Largest & $5^{th}$ Largest & & Sum \\ 
 \hline 
& \mbox{0 - 750} & & & $0.460 \pm 0.179$ & $0.212 \pm 0.081$ & $0.118 \pm 0.056$ & $0.071 \pm 0.040$ & $0.045 \pm 0.030$ & & $0.905$ \\
 \mbox{United States} & \mbox{750 - 1250} & mi & & $0.397 \pm 0.149$ & $0.208 \pm 0.072$ & $0.129 \pm 0.048$ & $0.084 \pm 0.038$ & $0.055 \pm 0.030$ & & $0.873$ \\
& \mbox{1250+} & & & $0.382 \pm 0.153$ & $0.201 \pm 0.061$ & $0.132 \pm 0.048$ & $0.083 \pm 0.036$ & $0.058 \pm 0.030$ & & $0.856$ \\
\hline 
& \mbox{0 - 150} & & & $0.396 \pm 0.148$ & $0.088 \pm 0.042$ & $0.061 \pm 0.030$ & $0.061 \pm 0.030$ & $0.046 \pm 0.024$ & & $0.739$ \\
\mbox{England} & \mbox{150 - 275} & mi & & $0.410 \pm 0.205$ & $0.169 \pm 0.087$ & $0.084 \pm 0.043$ & $0.057 \pm 0.031$ & $0.042 \pm 0.024$ & & $0.761$ \\
& \mbox{275+} & & & $0.390 \pm 0.014$ & $0.158 \pm 0.062$ & $0.097 \pm 0.033$ & $0.063 \pm 0.021$ & $0.044 \pm 0.018$ & & $0.752$ \\
\hline 
& \mbox{0 - 90} & & & $0.472 \pm 0.183$ & $0.199 \pm 0.085$ & $0.098 \pm 0.040$ & $0.070 \pm 0.029$ & $0.052 \pm 0.027$ & & $0.891$ \\
\mbox{Denmark} &\mbox{90 - 185} & mi & & $0.672 \pm 0.179$ & $0.139 \pm 0.088$ & $0.066 \pm 0.051$ & $0.037 \pm 0.026$ & $0.025 \pm 0.018$  & & $0.939$ \\
& \mbox{185+} & & & $0.564 \pm 0.100$ & $0.253 \pm 0.118$ & $0.064 \pm 0.040$ & $0.034 \pm 0.021$ & $0.023 \pm 0.014$ & & $0.937$ \\
\end{tabular}
\end{ruledtabular}
\caption{The average fraction $A_{j} = s_{j} / \ell $ of the total length covered by each of the five largest steps with standard deviations.}
\label{table:meanSteps}
\end{table*}

\begin{figure*}[t] 
\begin{center}
\begin{tabular*}{17.5cm}{ccc}
\includegraphics[scale=0.315]{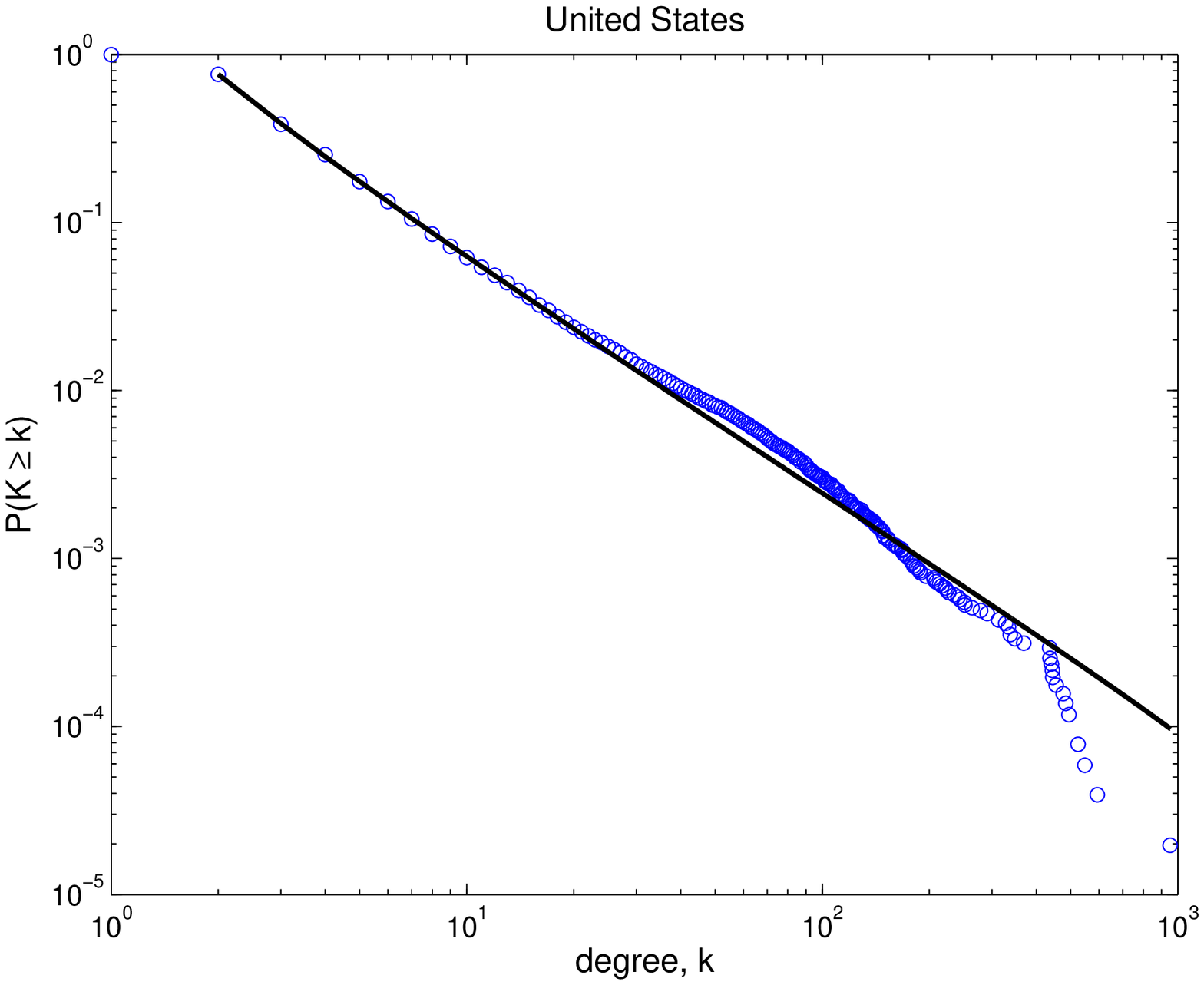} &
\includegraphics[scale=0.315]{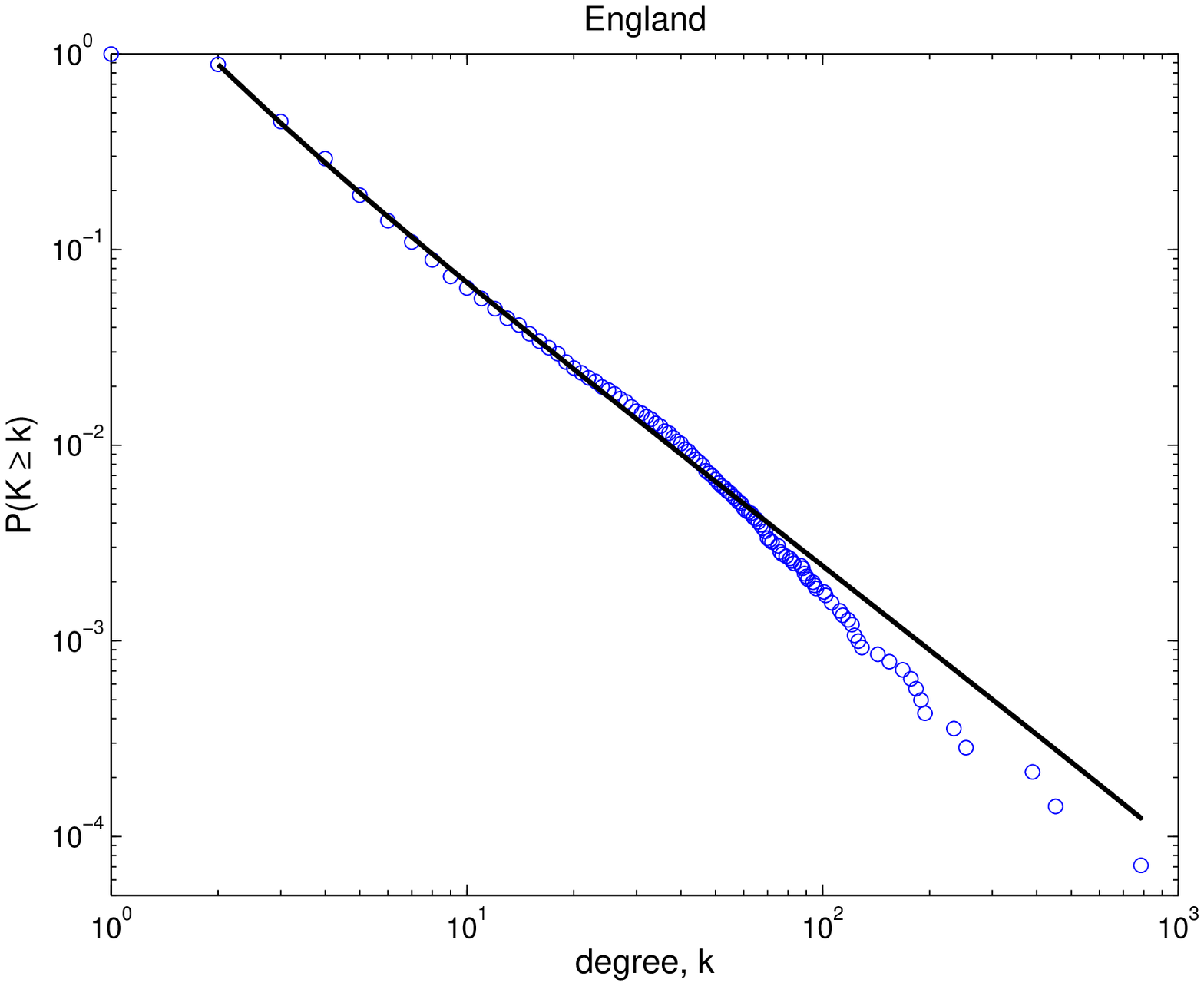} &
\includegraphics[scale=0.315]{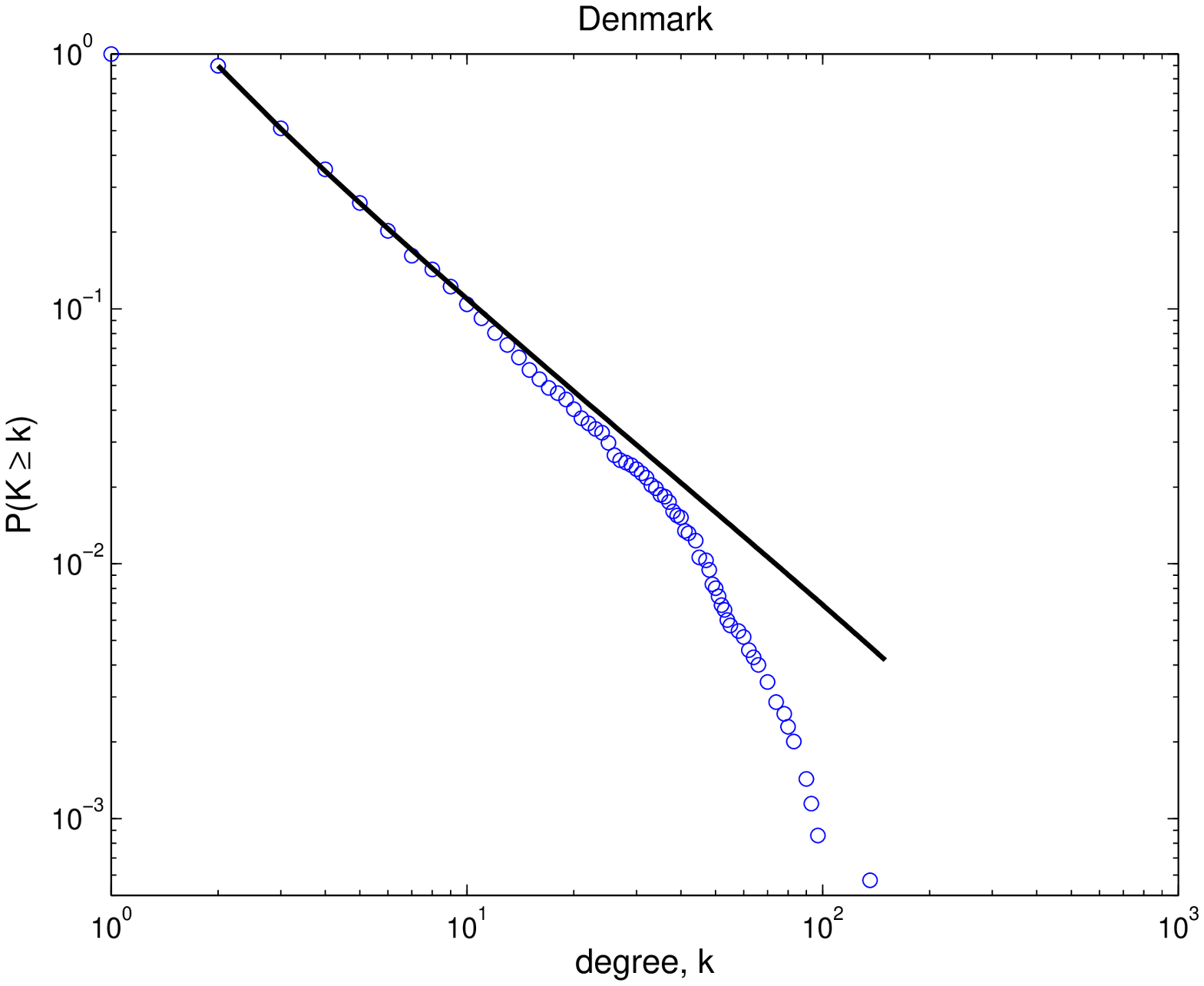} 
\end{tabular*}
\end{center}
\caption{(Color online) The cumulative degree distributions $P(K \geq k)$ of the dual model for the United States, England, and Denmark. We show fits based on a maximum likelihood estimate of a power law model $P(k) \sim k^{-\alpha}$, where $\alpha = 2.4, 2.2$ and $2.4$, respectively. }
\label{fig:degrees}
\end{figure*}

\section{Degree Distribution}
Other studies of road networks have found that the degree distribution of the dual graph, i.e., the number of intersections in which a single road is involved, is heavy-tailed, although not necessarily a power law~\cite{Jiang04, Rosvall05}. We find similarly heavy-tailed distributions at the national level (Figure~\ref{fig:degrees}), but with apparent finite-size cutoffs related to respective geographic scales. For small countries like England and Denmark, and for cities as in~\cite{Rosvall05}, the case for scale invariance is not clear: the data spans too small a range to rule out other heavy-tailed distributions like the log normal. Based on the nearly three decades of relatively clean scaling for the United States, however, we conjecture that the formation of road networks, when conducted at a sufficiently large scale, leads to true scale-free structure.



We fit these distributions using a maximum likelihood estimator for the power law, as in~\cite{Clauset05}. For the United States, England and Denmark, we find scaling exponents of $\alpha = 2.4, 2.2$ and $2.4$, respectively. These are likely slight overestimates of their true values; progressive sub-sampling of each network yields slightly larger estimates of $\alpha$. Using Monte Carlo simulation and the Kolmogorov-Smirnov goondess-of-fit test, in the manner of~\cite{Goldstein04}, we find that the models are good representations of the data, with $p_{MC} \geq 0.99$. Finally, we do not propose that $\alpha$ has a universal value. Rather, in the next section, we describe a toy model which can give rise to a variety of exponents, depending on the fractal dimensions describing the placement of roads and intersections.

\section{A Simple Fractal Model}
In this section we introduce and analyze a simple fractal model for the placement of roads on the unit square that reproduces both the observed hierarchical and scale invariant structure of journeys. As we will see, the key quantities of the model are the fractal or Hausdorff dimensions $d_{p}$ and $d_{i}$ that, in turn, describe the distribution of road intersections in the plane, and the distribution of intersections along a single road.

Unlike previous models of physical networks~\cite{Newman04, FKP02, Manna03}, our model assumes no optimization or resource constraint satisfaction mechanism. Rather, we simply assume the fractal structure is given, and analyze the resulting implications for journey structure and the dual degree distribution. We leave for future work the exploration of mechanisms that may in turn generate a fractal placement of roads.

\begin{figure}[t] 
\begin{center}
\begin{tabular}{cc} 
\includegraphics[scale=0.65]{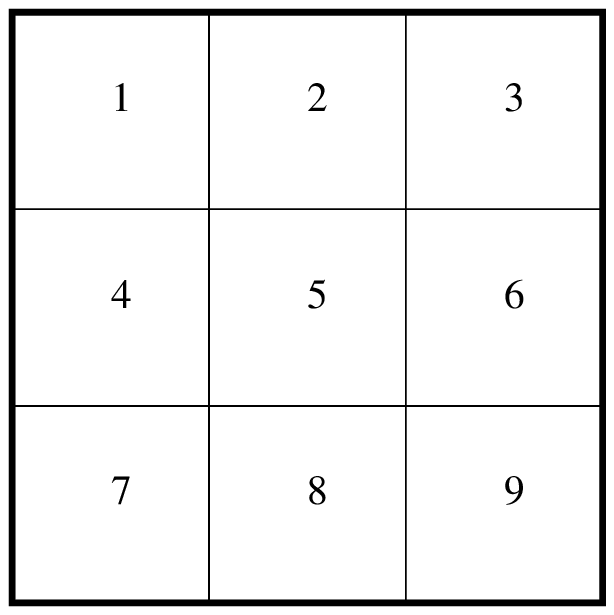} &
\includegraphics[scale=0.44]{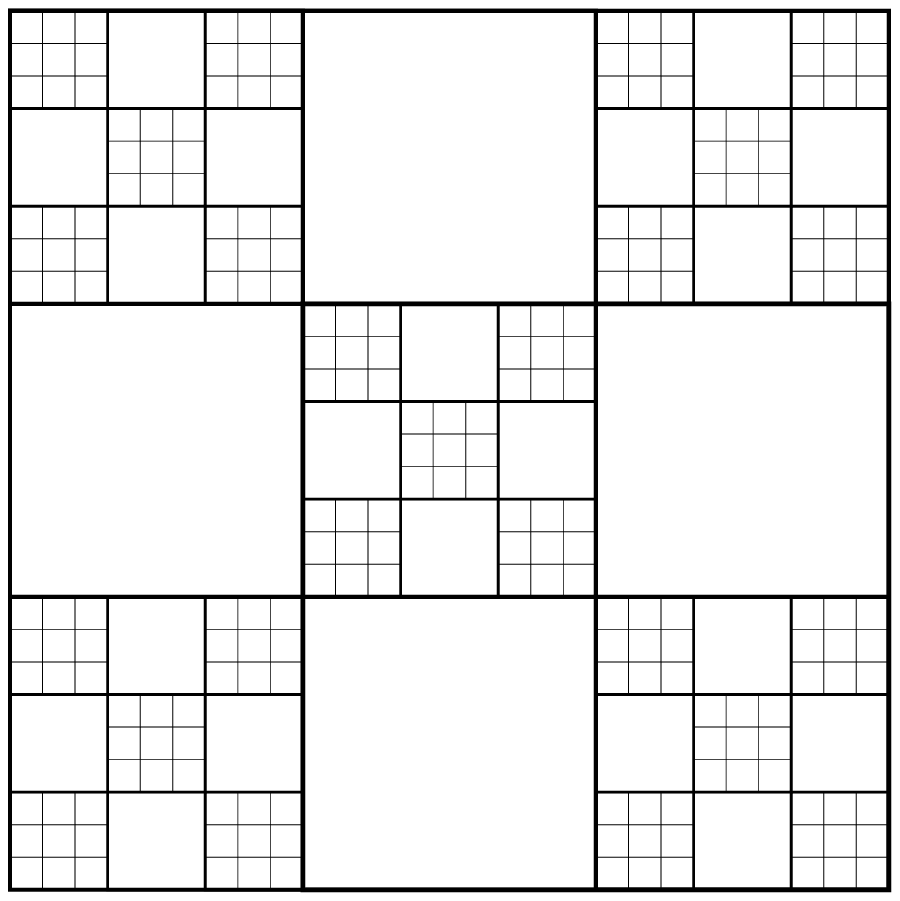} \\
\end{tabular}
\end{center}
\caption{A version of our fractal model for road placement (see text). Line-thickness indicates greater road capacity and speed limits. The Sierpinski carpet corresponds to recursively subdividing all squares except square 5.}
\label{fig:fractal}
\end{figure}

To create a road network according to our model, we first divide the unit square into $\eta^{2}$ squares of equal size for some fixed integer $\eta$ by placing $2(\eta-1)$ roads. We then choose some subset of these $\eta^{2}$ squares and subdivide them as we did the original square, by placing $2(\eta-1)$ new roads per subdivision. Repeating this process recursively for as many levels as desired yields a road network with fractal structure, where lines are roads and line-crossings are intersections. For instance, with $\eta=3$, subdividing all but the center square gives the Sierpinski carpet~\cite{SchroederM}, and in Figure~\ref{fig:fractal} we show a network resulting from subdividing five of the nine squares.

Observe that in this model the road intersections are distributed as a fractal both over the original unit square and along a given road.  For instance, in the Sierpinski carpet, at each level of construction the total number of intersections in the plane increases by a factor of $8$, while the number of intersections along a given road triples; this construction thus yields a fractal dimension \mbox{$d_{p} = \log_{3} 8$} for the distribution in the plane, and \mbox{$d_{i} = \log_{3} 3 = 1$} for the distribution along a given road. Similarly, for the scheme illustrated in Figure~\ref{fig:fractal}, at each level the number of intersections increases by a factor of $5$ and the number of intersections along a given road doubles, giving $d_{p} = \log_{3} 5$ and $d_{i} = \log_{3} 2$.

We show, by a simple counting argument, that the scaling exponent of such a network's dual degree distribution is related to the fractal dimensions in the following way,
\begin{equation}
\alpha = 1 + \frac{d_{p}}{d_{i}} \enspace .
\label{eqn:frac}
\end{equation}
For each $x\leq m$, where $m$ is the total number of levels of subdivision, the number of roads at level $x$ is 
$r(x) \sim \eta^{d_{p} \,x}$. Similarly, for a road added at level $x$, the number of intersections along its length is exponential in the number of subsequent subdivisions, and is given by $c(x) \sim \eta^{d_{i}(m-x)} \sim \eta^{-d_{i}\, x}$.


The cumulative degree distribution of this model can be calculated as follows. The number of roads with degree greater than $k$ is given by
\begin{align*}
P(K>k) & = \sum_{x~:~c(x) > k} r(x) \\
& \sim \sum_{x\,=\,m-(1/d_{i}) \log_{\eta} k}^{m} \,\eta^{d_{p}\, x} \\
& \sim k^{-d_{p}/d_{i}} \enspace .
\end{align*}
So, differentiating this cumulative distribution gives the degree distribution $P(k) \sim k^{-\alpha}$, with $\alpha$ given by Eq.~(\ref{eqn:frac}). The values of $\alpha$ for a few variations on the $\eta=3$ subdivision schema are given in Table~\ref{table:pls}.

\begin{table}[b]
\begin{center}
\begin{tabular}{|l|ccccc|}
\hline
Schema & $d_{p}$& & $d_{i}$&  & $\alpha$ \\
\hline
all    & $\log_3 9$ && $\log_3 3$ && $3.00$   \\
all but center   &  $\log_3 8$ && $\log_3 3$ && $2.89$   \\
odd numbers  &  $\log_3 5$ && $\log_3 2$ && $3.32$  \\
corners &$\log_3 4$ && $\log_3 2$ && $3.00$ \\
\hline
\end{tabular}
\end{center}\caption{Fractal dimensions for the distribution of intersections in the plane $d_{p}$, the distribution of intersections along a single road $d_{i}$, and the power-law exponent $\alpha$ for different subdivision schemes given by Eq.~\ref{eqn:frac} for $\eta=3$ (see text).}
\label{table:pls}
\end{table}

Further, by placing roads hierarchically through the subdivision process, journeys that seek to minimize travel time will necessarily utilize this same hierarchical structure, espeically if roads at the earlier levels of construction correspond to roads with higher speed limits and traffic capacities. For instance, if the source and destination are in different subsquares, then the shortest path in the dual model will use one of the roads at level $x=1$; this is also recursively true at each step of the journey. Thus, the $j$th largest step will cover an average fraction $A_{j}$ of the journey, which scales as $A_{j} \sim n^{-j}$. Indeed, looking at the data for the United States (Table~\ref{table:meanSteps}), it appears that $A_{j}$ decreases roughly exponentially with $j$.

The fact that our toy fractal model reproduces the scale invariant journey structure, and can similarly produce the correct functional form of the dual degree distribution, suggests that the roads in our real world networks may be organized in a similar fractal structure. It would be interesting to use the geographic distribution of population and road intersections to estimate the fractal dimensions $d_{p}$ and $d_{i}$ for various countries, and compare the value of $\alpha$ predicted by Eq.~(\ref{eqn:frac}) to the measured value. We leave this as a direction for future work.

\section{Conclusion}
We studied the national road networks of the United States, England and Denmark through their dual representation, using the driving directions provided by a popular commercial service. Like those of urban road networks~\cite{Rosvall05}, we found that the dual degree distribution is characterized by a  heavy tail; however, for large countries such as the United States, this distribution is likely scale-free, following a power law of the form $P(k)\sim k^{-\alpha}$, with $2.2 \leq \alpha \leq 2.4$ (Fig.~\ref{fig:degrees}).

We further showed that journeys on these networks have a scale invariant structure, characterized by a driver rising up through the road hierarchy, i.e., from local to regional to national roads where the speed limit and capacity grows with each step, and then descending in reverse order as she approaches her destination. This scale invariance is exhibited by the fact that journeys have similar structure regardless of their total length. Notably, our empirical observations say nothing about the actual traffic density on these roads, which is likely determined by the non-uniform popularity of destinations.

To explain the observed structure in the road networks, we introduced and analyzed a simple fractal model of road placement. This model recovers the scale-free structure of journeys in the network and the power-law dual degree distribution. It also suggests a fundamental relationship between the exponent $\alpha$ and the fractal dimensions describing the distribution of road intersections in the plane $d_{p}$ and along a single road $d_{i}$. Although our model assumes that road placement is not a function of resource-bound optimization as in~\cite{FKP02, Manna03}, it would be interesting to adapt it in such a way as to generate more statistically realistic road networks. Arguably, biological transportation networks, e.g., vascular networks, also have a fractal structure~\cite{West99}, and a comparative study of these and our road networks would be interesting.

\acknowledgements{
The authors thank Sahar Abubucker for reviewing the manuscript and Vincent Nesme for conversations on Hausdorff dimensions. This work was supported by NSF grants PHY-0071139, PHY-0200909, and the Sandia University Research Project. }

\newpage


\begin{thebibliography}{99}
\bibitem{footnote}
We note that while not everyone navigates in this way, this is the form of the results returned by the increasingly popular navigational services from companies such as Google, Yahoo!, Microsoft and Mapquest.
\end{thebibliography}
\end{document}